\documentstyle[aps,epsfig,multicol,graphicx]{revtex}
\def\breakon{\end{multicols}\widetext\vspace{-.2cm}
\noindent\rule{.48\linewidth}{.3mm}\rule{.3mm}{.3cm}\vspace{.0cm}}

\renewcommand{\vec}[1]{{\mathbf #1}}

\def\breakoff{\vspace{-.2cm}
\noindent
\rule{.52\linewidth}{.0mm}\rule[-.27cm]{.3mm}{.3cm}\rule{.48\linewidth}{.3mm}
\vspace{-.3cm}
\begin{multicols}{2}
\narrowtext}

\newcommand{\be}{\begin{equation}}
\newcommand{\ee}{\end{equation}}
\newcommand{\bea}{\begin{eqnarray}}
\newcommand{\eea}{\end{eqnarray}}

\newcommand{\p}{\partial}

\newcommand{\lb}{\left[}
\newcommand{\rb}{\right]}
\newcommand{\lp}{\left(}
\newcommand{\rp}{\right)}
\renewcommand{\phi}{\varphi}
\renewcommand{\vec}[1]{{\bf #1}}

\begin{document}

\title{Angular distribution of photoluminescence as a probe 
of Bose Condensation of trapped excitons}
\author{Jonathan Keeling$^1$, L.S. Levitov$^2$, P.B. Littlewood$^{1,3}$}
\address{$^1$Cavendish Laboratory, Madingley Road, Cambridge CB3 OHE, U.K.}
\address{$^2$Department of Physics,
Center for Materials Sciences \& Engineering,
Massachusetts Institute of Technology, 77 Massachusetts Ave,
Cambridge, MA 02139}
\address{$^3$National High Magnetic Field Laboratory, Pulsed Field Facility, 
LANL, Los Alamos NM 87545}

\maketitle
\begin{abstract}
Recent experiments on two-dimensional exciton systems have shown
the excitons collect in shallow in-plane traps. We find that
Bose condensation in a trap results in a dramatic change of the exciton
photoluminescence (PL) angular distribution. The long-range coherence 
of the condensed state gives rise to a sharply focussed 
peak of radiation in the direction normal to the plane. 
By comparing the PL
profile with and without Bose Condensation we provide a simple
diagnostic for the existence of a Bose condensate.
The PL peak has strong temperature dependence due to the  thermal
order parameter phase fluctuations across the system.
The angular PL distribution can also be used for imaging vortices in the 
trapped condensate. Vortex phase spatial variation 
leads to destructive interference of PL radiation in certain directions, 
creating nodes in the PL distribution that imprint the vortex configuration.
\end{abstract}


\begin{multicols}{2}

\narrowtext
The possibility of Bose condensation in two-dimensional exciton systems,
such as coupled quantum wells, 
has been actively investigated recently 
\cite{Butov01,Timofeev,Butov02traps,Butov02,Snoke02,SnokeScience}. 
The interest in these systems is caused by relatively long recombination times
that should make it possible to create a sufficiently cold system
at high density, and thereby reach the condensation transition point. 
Further, their two dimensional nature leads to a repulsive interaction,
disfavouring the formation of biexcitons.
It was proposed that
the condensation of excitons 
can be facilitated in the presence of
in-plane traps \cite{Butov94Lum}.
Since the temperature of the exciton gas depends on the intensity
of the laser used to pump excitons, it is advantageous if the
excitons can reach critical density at lower pumping power.  By
collecting excitons in a trap, a lower total flux of excitons and
hence a lower pumping power is required. 
Small traps of micron size arising due to ambient disorder 
\cite{Butov02traps} (see also \cite{Butov03}), as well as
more shallow traps {\it ca.}\ hundred microns wide created artificially 
by stress \cite{Snoke99traps} have been considered. 

The density distribution of excitons in such relatively shallow traps 
(of the order of $1\,{\rm meV/\mu m}$)
changes 
very little through the condensation transition. Thus, in contrast with 
the cold atom systems\cite{Cornell,Ketterle}, 
the direct spatial imaging of density is not expected 
to provide dramatic evidence for condensation. The principal manifestations 
of Bose condensation discussed to date include changes in PL spectrum 
\cite{Timofeev,Fukuzawa}
and exciton recombination rate \cite{Butov01,Butov94Lum}. 

In this work, we demonstrate that PL angular distribution can be used 
as a sensitive probe of Bose condensation in a trap. 
For excitons confined to move in a plane, in recombination
momentum conservation perpendicular to this plane is relaxed.  The
intensity of radiation at a particular angle is therefore controlled by
the distribution of in plane momenta of the excitons. 
Bose condensation produces large occupation 
of the state with zero velocity, thereby creating a peak 
in the momentum distribution, which gives rise to 
a sharp peak in the PL angular profile. Well below the transition, 
once the long-range order
in the condensate extends across the entire exciton cloud in trap,
the phases of the condensate order parameter become correlated throughout 
the cloud. This makes the PL radiation from different parts of the cloud
fully coherent, and thus focussed in a cone with opening angle
{\it ca.} $\lambda_{\rm rad}/L$, 
with $\lambda_{\rm rad}$ the optical wavelength 
and $L$ the cloud size.

To extract the PL angular profile, we consider dipole radiation of excitons, 
with momentum
conservation perpendicular to the plane relaxed
\cite{andreani91:_radiat_lifet,chuang95:_physic_optoel},
\begin{eqnarray}
  I(\vec k_\parallel ) \propto
  \label{eq:1}
  N_{\rm ex}(\vec k_\parallel )
  \rho_{\rm ph}(\vec{k}_\parallel , k_0)
  \left|
    \left< f\right|
    \vec{k}_\parallel.\hat{\vec{p}}
    \left|i\right>
  \right|^2 
\end{eqnarray}
where $k_0=2\pi/\lambda_{\rm rad}$ is photon wavenumber, 
$N_{\rm ex}(\vec{k}_\parallel )$ is the number of excitons with
momentum $\vec{k}_\parallel $, $\rho_{\rm ph}(\vec{k}_\parallel , k_0)$ 
the photon density of
states, and $\left|i\right>$ and $\left|f\right>$ represent the initial
and final electronic states.
After $|\vec k_\parallel |$ is expressed in terms of the angle $\theta$ to the
plane normal, $|\vec k_\parallel| = k_0 \sin\theta$, the photon density of states and
matrix element give a factor $(1+\cos^2\theta)/\cos\theta$.

The momentum distribution of excitons depends on the state of the
Bose gas:
%
\be
  \label{eq:2}
  N_{\rm ex}(\vec{k}_\parallel ) 
  =
  \left<
|\psi(\vec{k}_\parallel )|^2 \right> 
  =
  \sum_{\vec{r},\vec{r}^{\prime}}
  \left<\psi^\ast(\vec{r})\psi(\vec{r}^{\prime}) \right> 
  e^{i \vec{k}_\parallel.(\vec{r}-\vec{r}^{\prime})}
\ee
%
We model the system as a gas of interacting bosons moving in
an external trap potential $V(r)$,
\begin{equation}
  \label{eq:3}
  H[\psi] =
  \int \lb
  \psi^*
  \left( - \frac{\nabla^2}{2m} + V(\vec r) - \mu\right)
  \psi +
  \frac{\lambda}{2} \left| \psi \right|^4\rb d^2\vec r
\end{equation}
where $\lambda$ is a momentum independent interaction.

In the Thomas-Fermi approximation \cite{PethickSmith}, 
with the system energy 
being a functional of local density only, one can look for
the minimum of the energy (\ref{eq:3}) with the gradient term omitted. For
an harmonic trap  $V(\vec r)=\frac12 \alpha \vec r^2$ containing $N$ particles, 
this gives a parabolic distribution
\begin{equation}
  \label{eq:4}
  \left|\psi(\vec r)\right|^2 = \frac{2N}{\pi R^4}(R^2 -\vec r^2)
\,,\quad
  R=\left({\frac{4 \lambda N}{\pi \alpha}}\right)^{1/4}
,
\end{equation}
while $\psi_{|\vec r|>R}=0$. The momentum space profile is
\begin{equation}
  \label{eq:5}
  N_{\rm ex}(\vec{k}) 
  =
8\pi R^2 N
  \left(
    \frac{\sin k R}{(k R)^3} - \frac{\cos k R}{(k R)^2}
  \right)^2
\end{equation}
(for brevity, from here on we use $\vec k\equiv\vec k_\parallel$).
Oscillations in the expression (\ref{eq:5}) result from the sharp edge
of the Thomas-Fermi distribution (\ref{eq:4}). 

\begin{figure}[t]
\centerline{%
\begin{minipage}[t]{3.5in}
\vspace{-10pt}
\hspace{-10pt}
\centering
\includegraphics[width=3.5in]{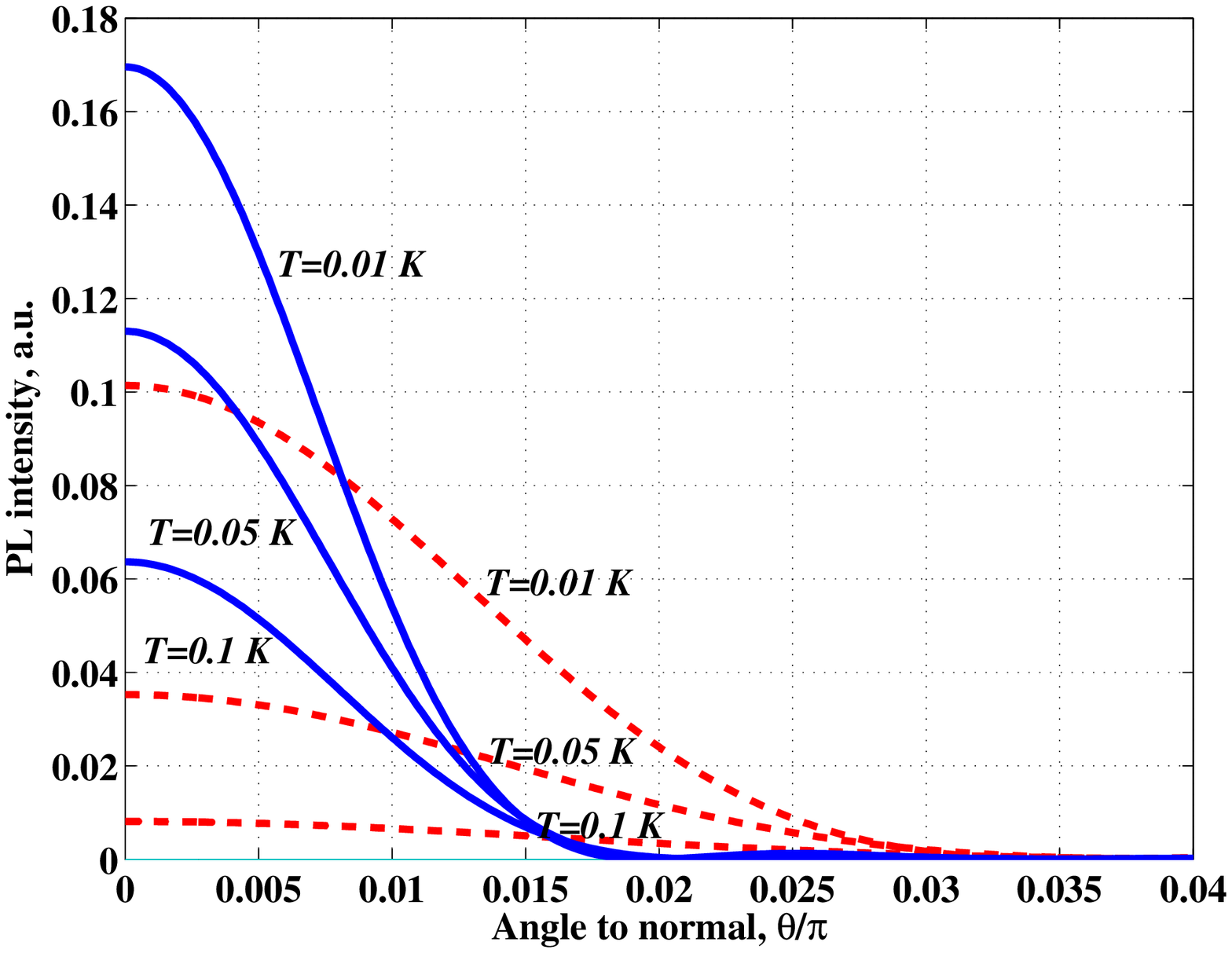}
\end{minipage}
\hspace{-1.8in}
\begin{minipage}[t]{1.4in}
\vspace{0.2in}
\centering
\includegraphics[width=1.4in]{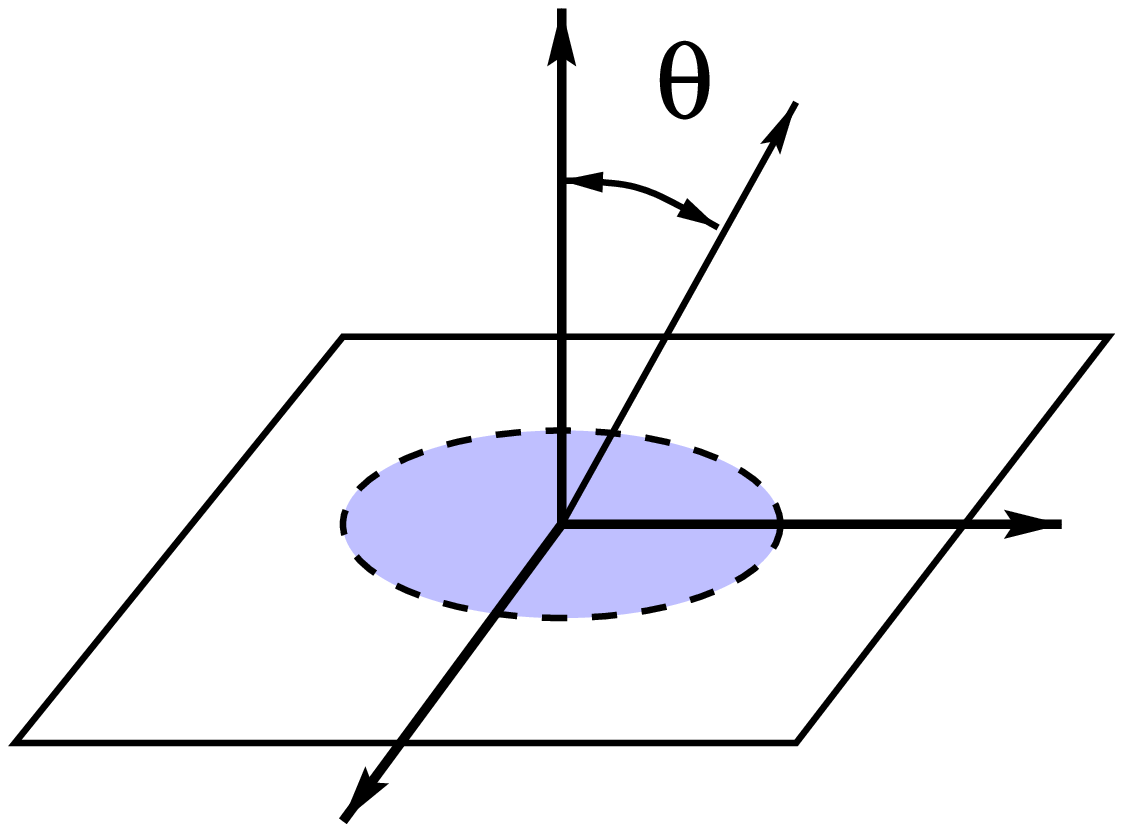}
\end{minipage}
}
\vspace{0.15cm}
\caption[]{%
PL angular profile temperature dependence for particle number in the trap
$N=10^3$ (solid lines) and $N=10^2$ (dashed lines).
Temperature dependence is calculated for
the parameters as discussed in the text, with
$T_{\rm BEC}$ given by Eq.(\ref{eq:Tbec}).
The intensity for $N=10^3$ particles is reduced by a factor $1/20$ 
to allow comparison.
Inset:
PL radiation geometry of a 2D exciton cloud.
}
\label{fig:peak}
\end{figure}

The momentum distribution (\ref{eq:5}) will produce a sharp peak 
in the PL angular profile if $R>\lambda_{\rm rad}$.  For the indirect
exciton line $\lambda_{\rm rad} \approx 800\,{\rm nm}$ \cite{Butov02traps}.  
By either considering the
capacitance of such a device, or the momentum independent part of the
Fourier transform of the interaction, we estimate that 
$\lambda = e^2 d/\epsilon
\approx 7.0 \,{\rm nm^2 eV}$ where $d \approx 5 \,{\rm nm}$
 is the inter-well spacing and $\epsilon$
the dielectric constant in GaAs.  

A characteristic scale for the trap potential is $\alpha \approx
10^{-12} \,{\rm eV\, nm^{-2}} $ (i.e. $\Delta V=0.1\,{\rm meV}$ 
for $\Delta r=10\,{\rm \mu m}$, \cite{Butov02traps}).
With these values, since the Thomas-Fermi radius
$R\sim 1.5\,N^{1/4}{\rm \mu m}$, 
a peak should be visible even for small particle numbers.

The PL radiation geometry schematic and the peak, 
along with its temperature 
dependence discussed below, are shown in Fig.\ref{fig:peak}.

The Bose condensation temperature can be estimated 
from the noninteracting model \cite{Butov02traps}, 
$T_{\rm BEC}=\hbar \sqrt{\alpha N/gm}$, 
with $g=4$ the exciton degeneracy factor.
Alternatively, since in a strongly interacting system
the density change for $T<T_{\rm BEC}$ is small,
$T_{\rm BEC}$ can be estimated from the degeneracy temperature at the 
density $\rho_{\rm max}$ of the centre of the Thomas-Fermi profile, giving
\be\label{eq:Tbec}
T_{\rm BEC} = \frac{2\pi\hbar^2}{m}\sqrt{\frac{N\alpha}{\lambda}}
\approx 0.02 \sqrt{N} \, [{\rm K}]
\ee
(for exciton mass $m=0.21\,m_e$, 
the value $m/2\pi \hbar^2 = 0.438\,{\rm nm^{-2}eV^{-1}}$). 
The two figures for $T_{\rm BEC}$ only differ by about a factor of $2$, because
$\sqrt{\lambda m/2\pi\hbar^2} \approx 2$. With $N=10^4$ we obtain 
$T_{\rm BEC}\approx 2\,{\rm K}$.  
(This corresponds to an exciton density of 
$\sim 10^{-4}\,{\rm nm}^{-2}$, comfortably within the dilute limit.)

\begin{figure}[t]
\centerline{%
\begin{minipage}[t]{0.9in}
\vspace{0.2pt}
\centering
\includegraphics[width=0.9in]{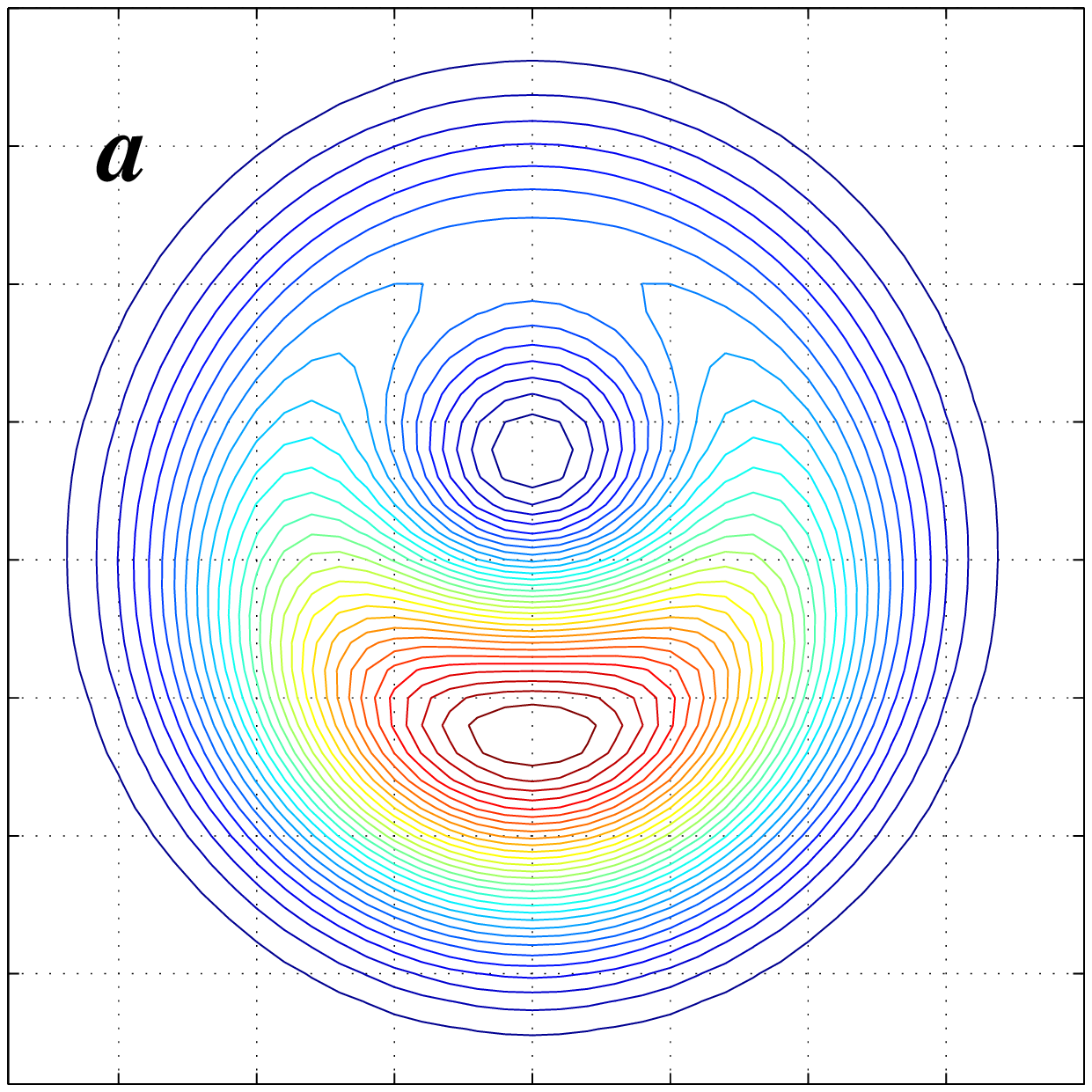}
\end{minipage}
\hspace{-0.1in}
\begin{minipage}[t]{0.9in}
\vspace{0.0in}
\centering
\includegraphics[width=0.9in]{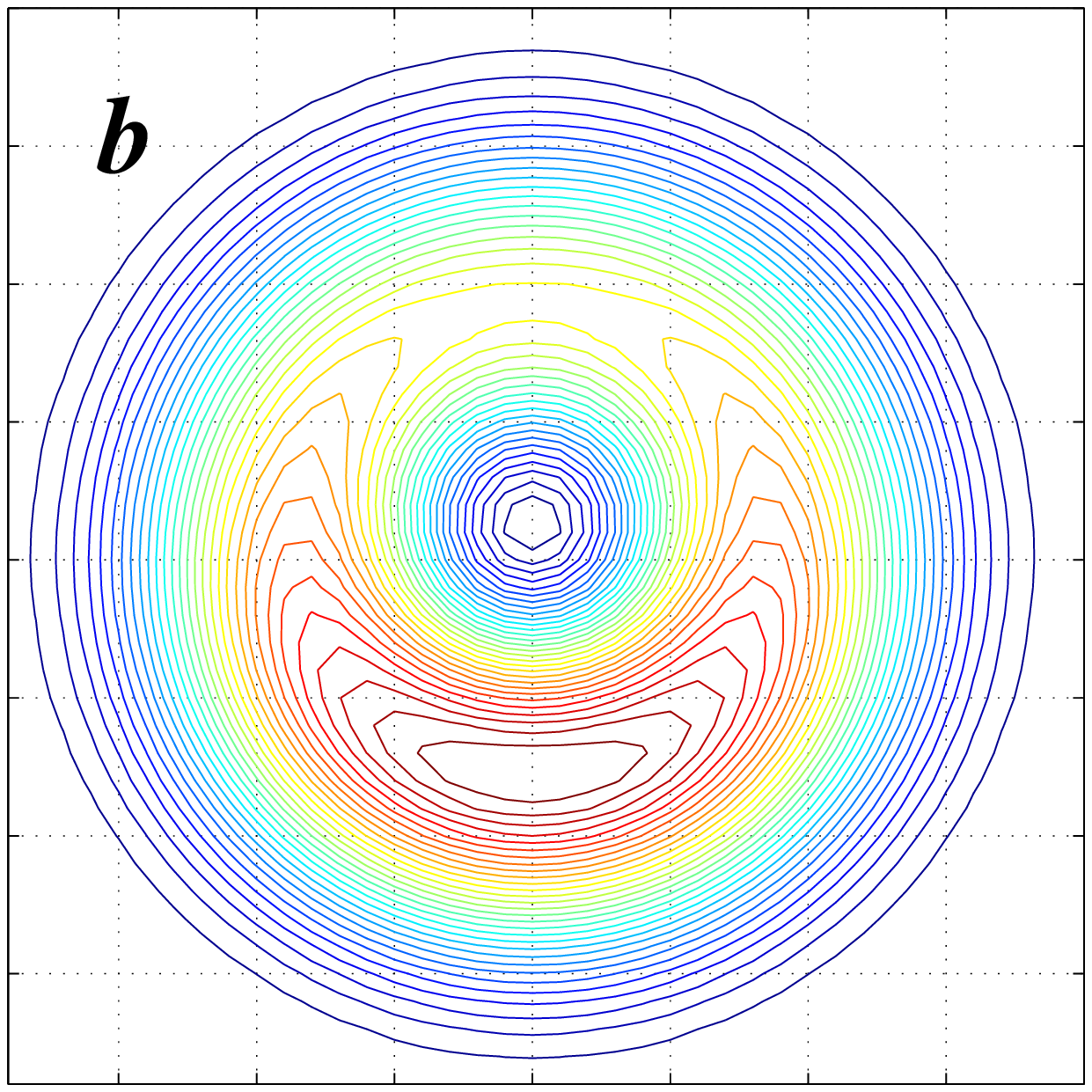}
\end{minipage}
\hspace{-0.1in}
\begin{minipage}[t]{0.9in}
\vspace{0.2pt}
\centering
\includegraphics[width=0.9in]{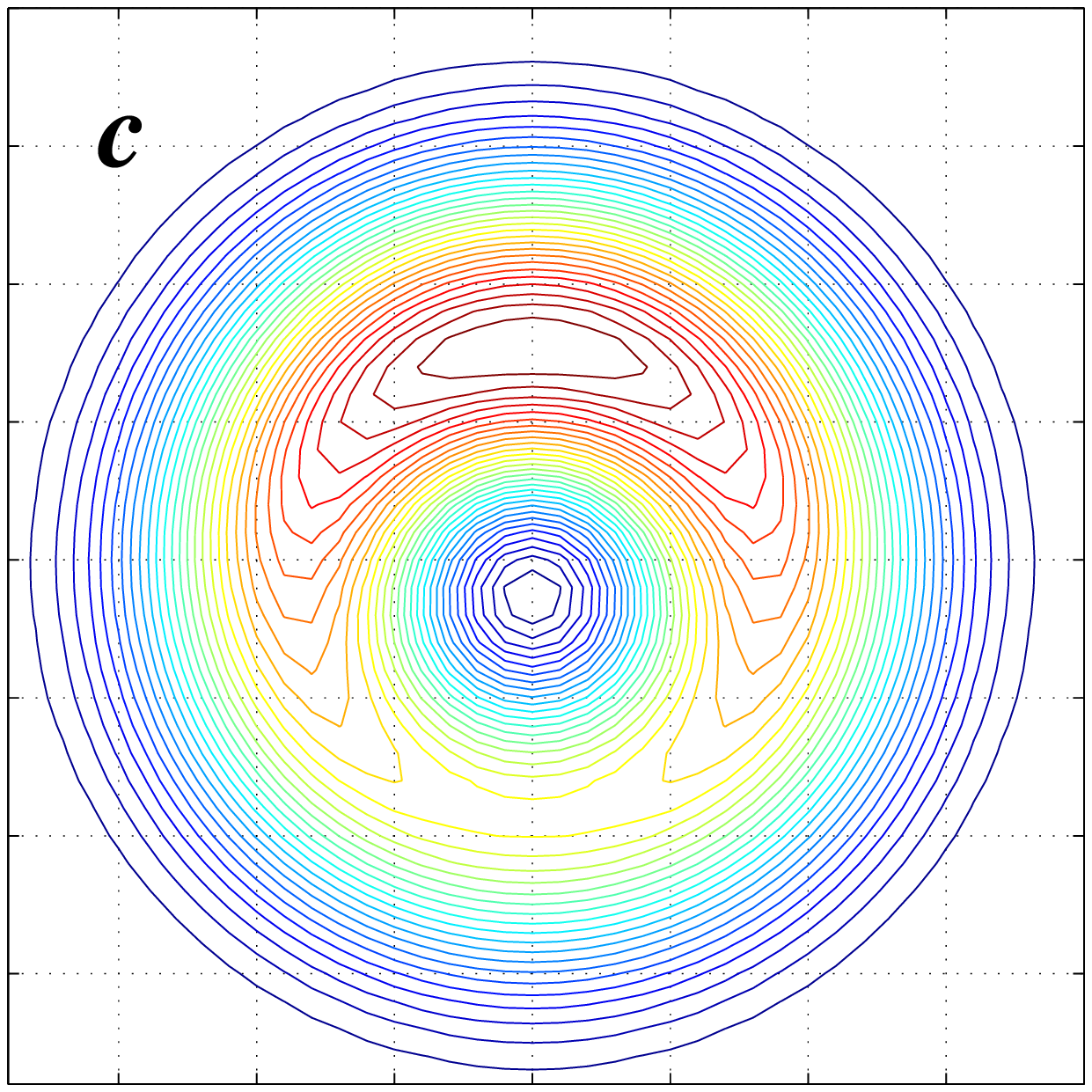}
\end{minipage}
\hspace{-0.1in}
\begin{minipage}[t]{0.9in}
\vspace{0.0in}
\centering
\includegraphics[width=0.9in]{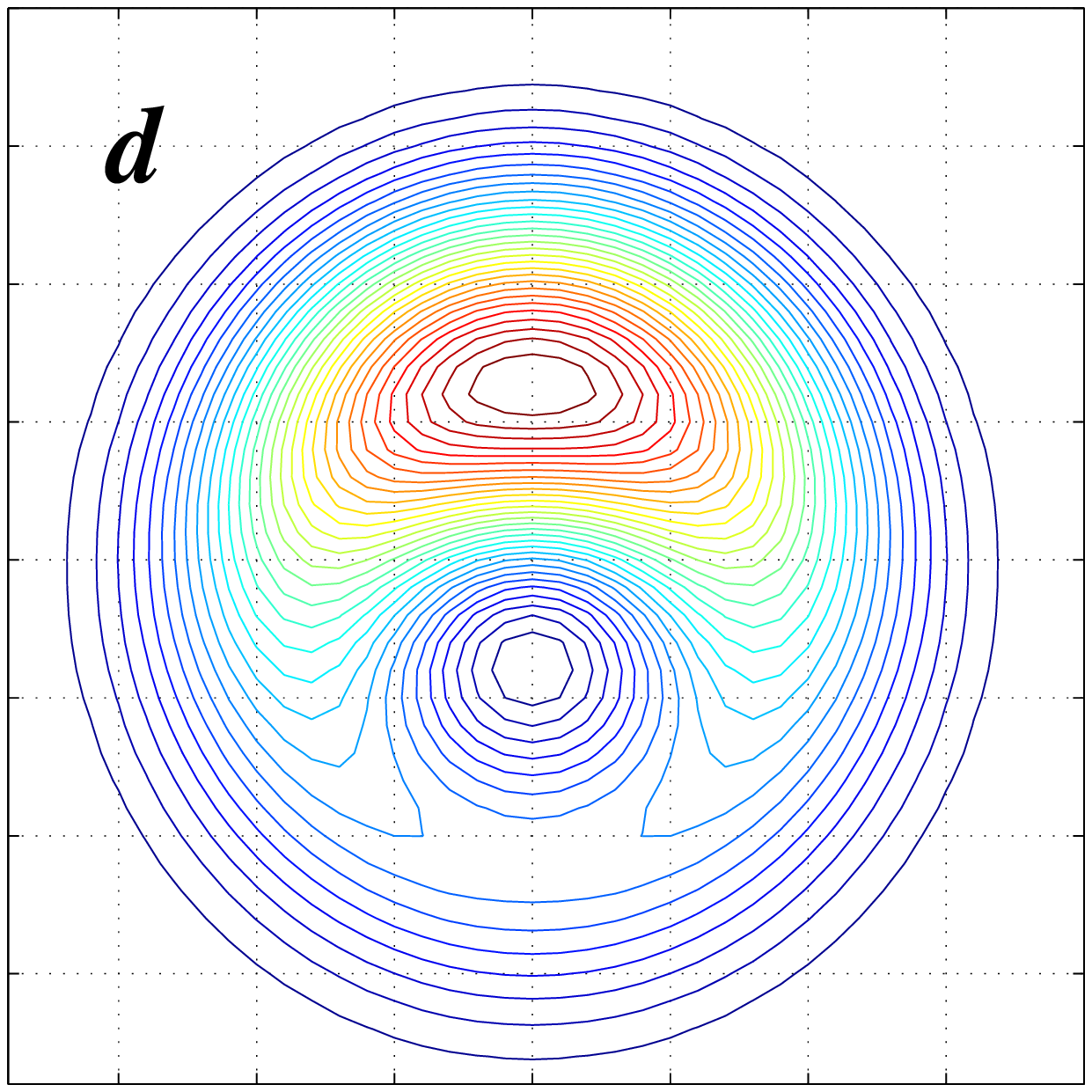}
\end{minipage}
}
\vspace{0.15cm}
\caption[]{%
The correspondence between vortices and nodes in PL angular profile. 
Left to right:
momentum distribution (\ref{eq:2}) for Thomas-Fermi condensate 
(\ref{eq:4})
with a single vortex
(\ref{eq:phase_vortices})
located on the $x$ axis at
$x/R=-0.3,\,-0.1,\,0.1,\,0.3$.
The off-center displacement of nodes in $\vec k$ space is $90^\circ$-rotated
with respect to vortex displacement in position space.
}
\label{fig:vort_move}
\end{figure}

The angular PL profile of the condensate 
changes in an interesting way in the presence of vortices. 
Here we study the momentum distribution for a condensate with $n$ vortices,
and show that the vortices are `imaged' by the nodes in $\vec k$-space. 
Consider the condensate (\ref{eq:4})
containing $n$ vortices. Using
the complex variable $z=x+iy$ in the plane to write the phase factors 
due to the vortices located at the points $z_j$, we have
\be\label{eq:phase_vortices}
\psi_n(\vec r)=\psi_0(\vec r)\exp\lp i\sum_{j=1}^n {\rm arg}(z-z_j)\rp
\ee
with $\psi_0$ defined by (\ref{eq:4}). 
Vortices give rise to nodes in the Fourier transform of
$\psi_n(\vec r)$ and thereby in the momentum distribution 
$N_{\rm ex}(\vec k)$ (\ref{eq:2}).
%
%
There is a one-to-one relation between the configuration of the nodes 
in $N_{\rm ex}(\vec k)$
and the vortex positions. Before discussing the more complicated 
case of phase vortices
(\ref{eq:phase_vortices}), it is instructive to look at an example of a
condensate function of the form
\be\label{eq:prod(z-zj)}
\psi_n(z)=A\prod_{j=1}^n(z-z_j)\exp\lp -\bar zz/2R^2\rp
\ee
describing $n$ `vortices' with both the amplitude and phase varying in space.
In this case the configuration of nodes in the momentum distribution
is a very simple one. 
Evaluating the Fourier transform, we represent the polynomial 
$\prod_{j=1}^n(z-z_j)$ as a differential operator, and obtain
\bea
\psi_n(\vec k)&=&2\pi A\prod_{j=1}^n(-2i\p_{\bar k}-z_j)\exp\lp -R^2\bar kk/2\rp
\nonumber\\
&=& 2\pi A\prod_{j=1}^n(i R^2 k-z_j)\exp\lp -R^2\bar kk/2\rp
\eea
with the complex variable $k=k_x+ik_y$.
This expression has $n$ nodes at the points
\be
k_j=-i z_j/R^2
\,,\quad j=1,\ldots,n,
\ee 
arranged in exactly the same way as vortices in real space,
albeit rotated by $90^{\circ}$ in a clockwise direction.

\begin{figure}[t]
\centerline{
\begin{minipage}[t]{1.2in}
\vspace{0.2pt}
\centering
\includegraphics[width=1.2in]{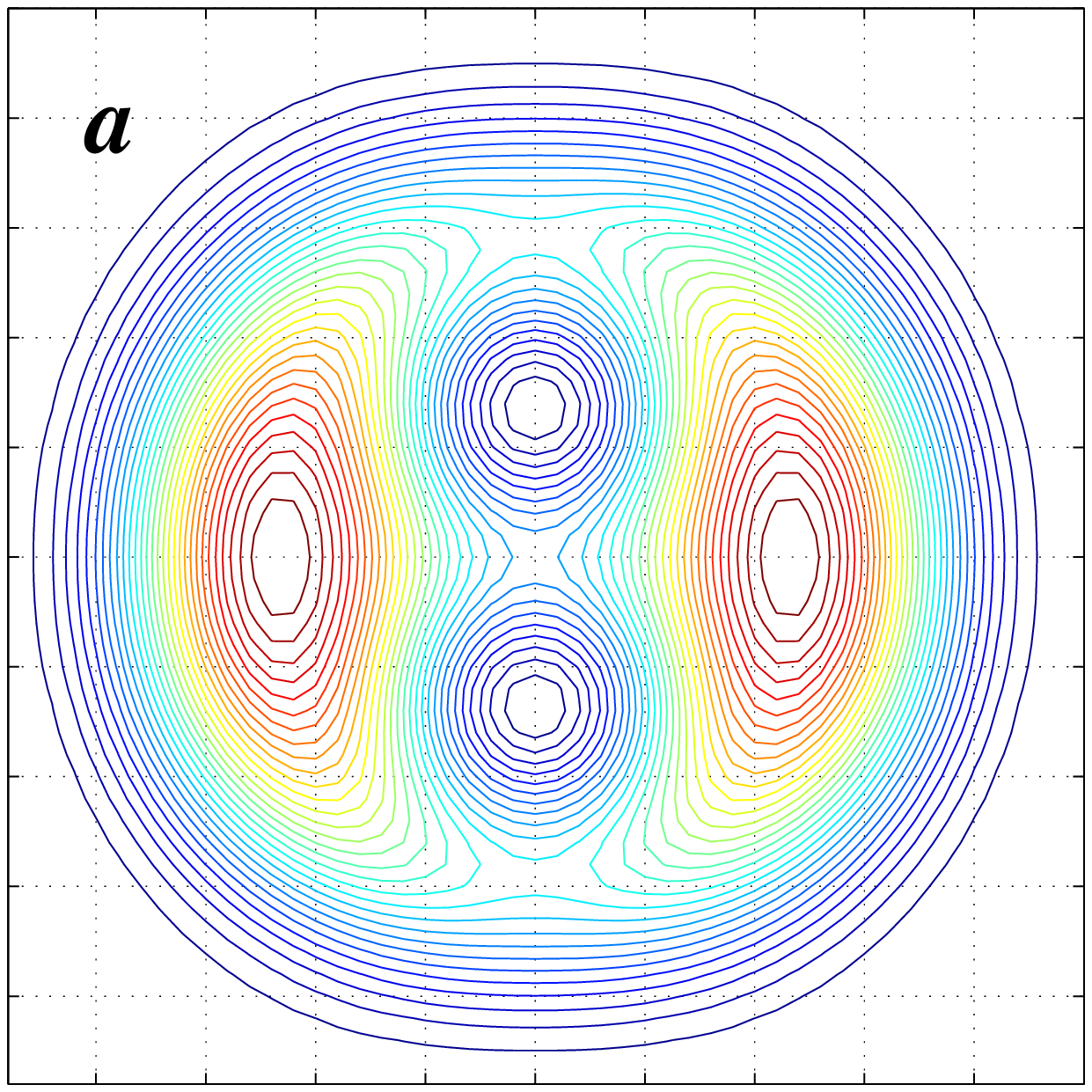}
\end{minipage}
\hspace{-0.5in}
\begin{minipage}[t]{0.4in}
\vspace{0.8in}
\includegraphics[width=0.4in]{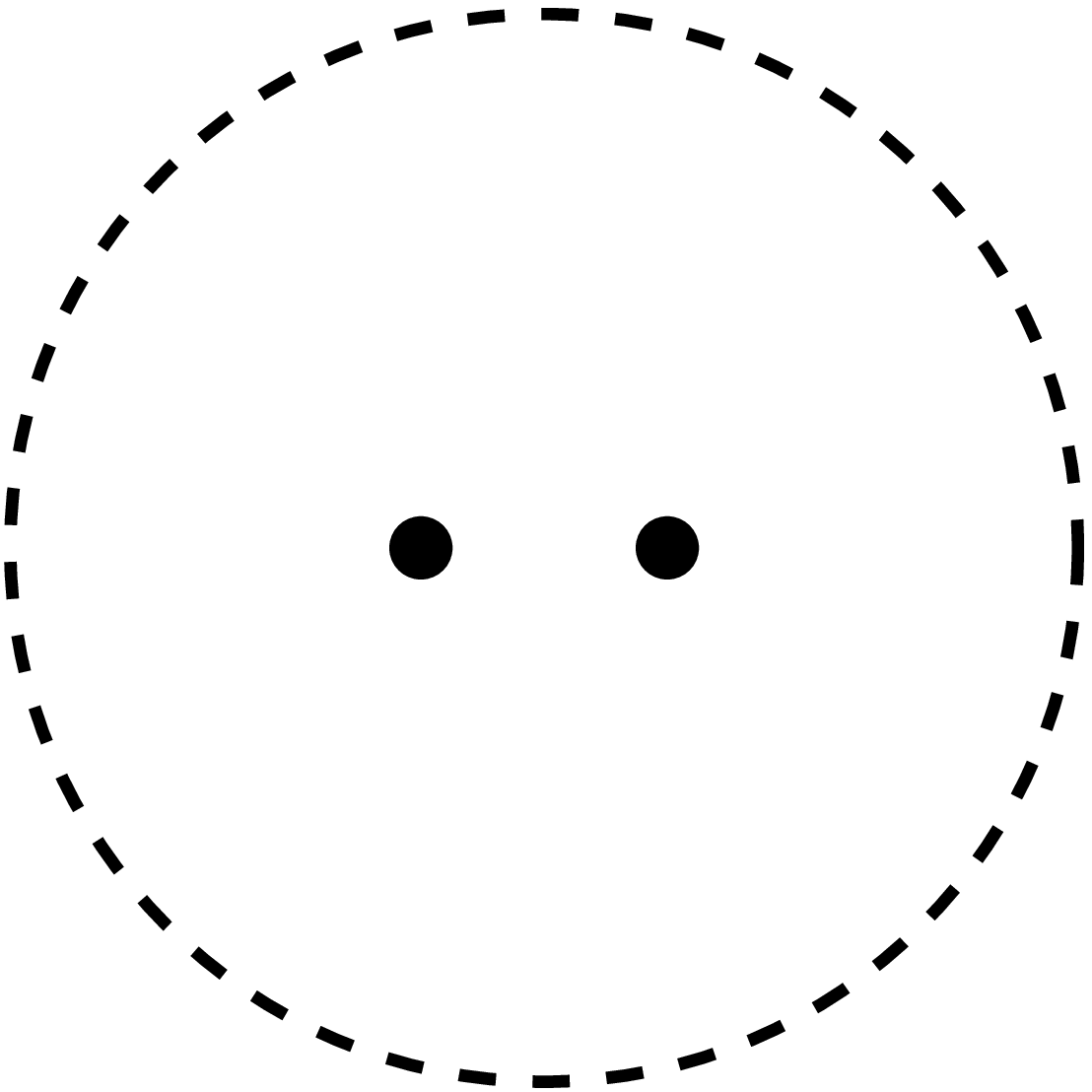}
\end{minipage}
\begin{minipage}[t]{1.2in}
\vspace{0.0in}
\centering
\includegraphics[width=1.2in]{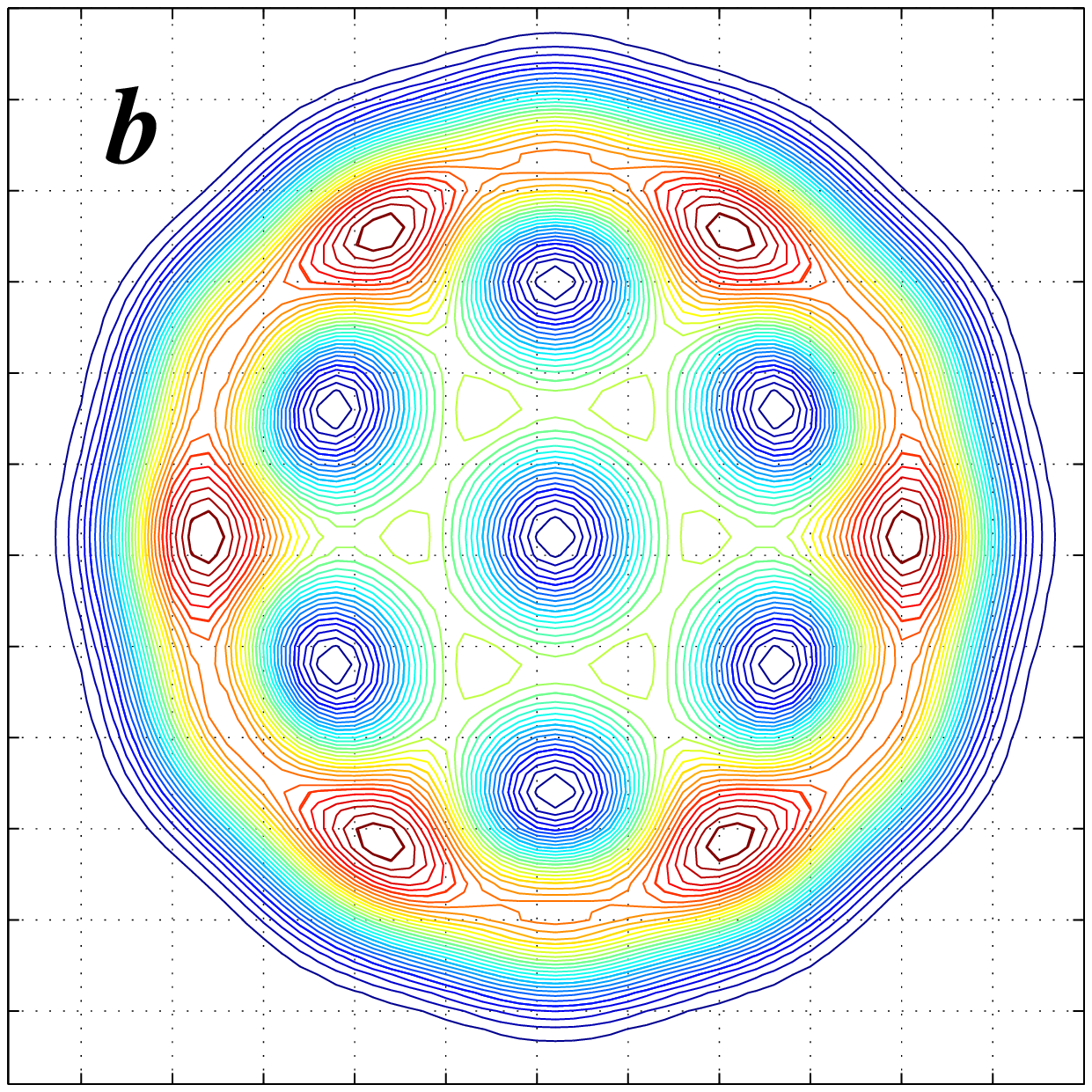}
\end{minipage}
\hspace{-0.5in}
\begin{minipage}[t]{0.4in}
\vspace{0.8in}
\includegraphics[width=0.4in]{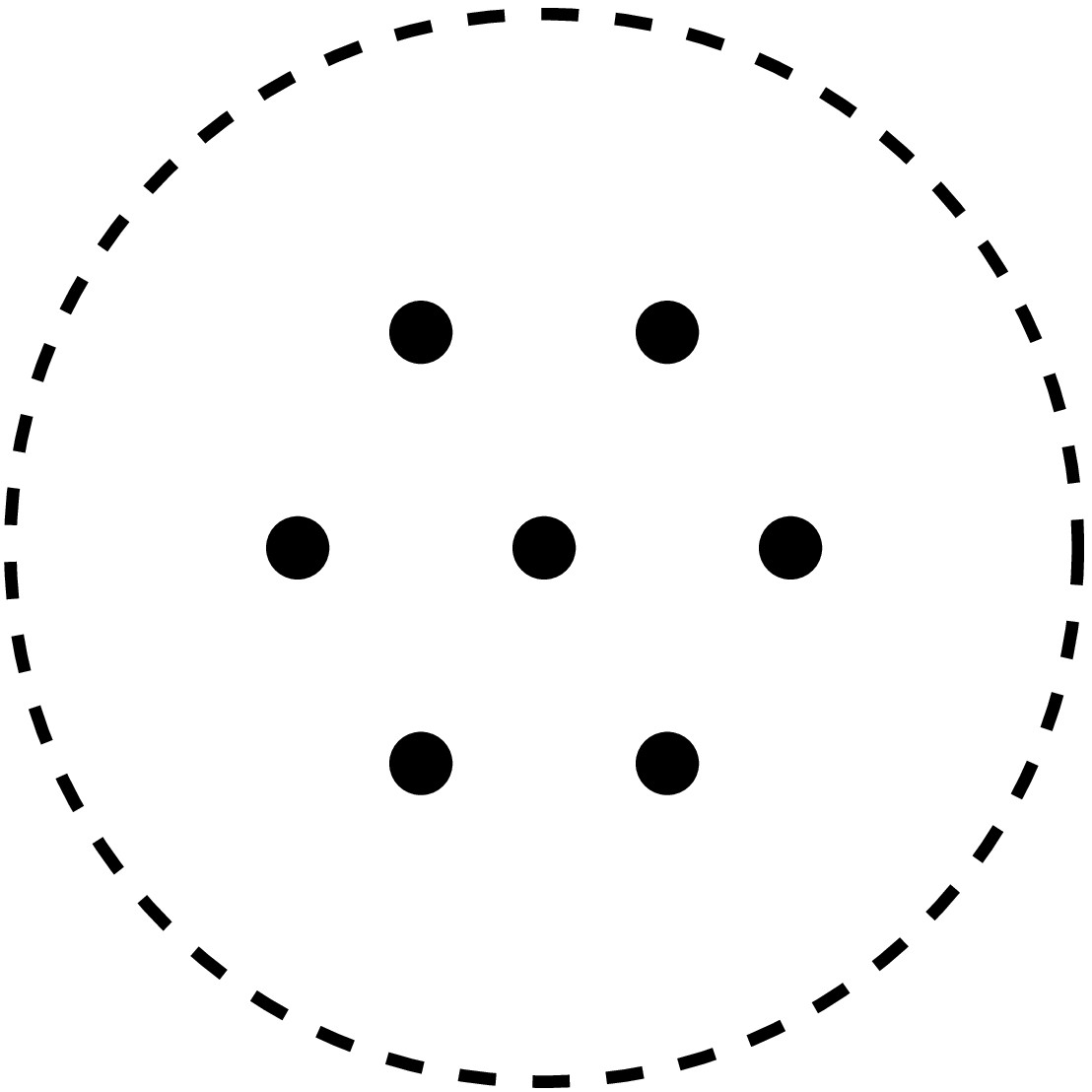}
\end{minipage}
\begin{minipage}[t]{1.2in}
\vspace{0.2pt}
\centering
\includegraphics[width=1.2in]{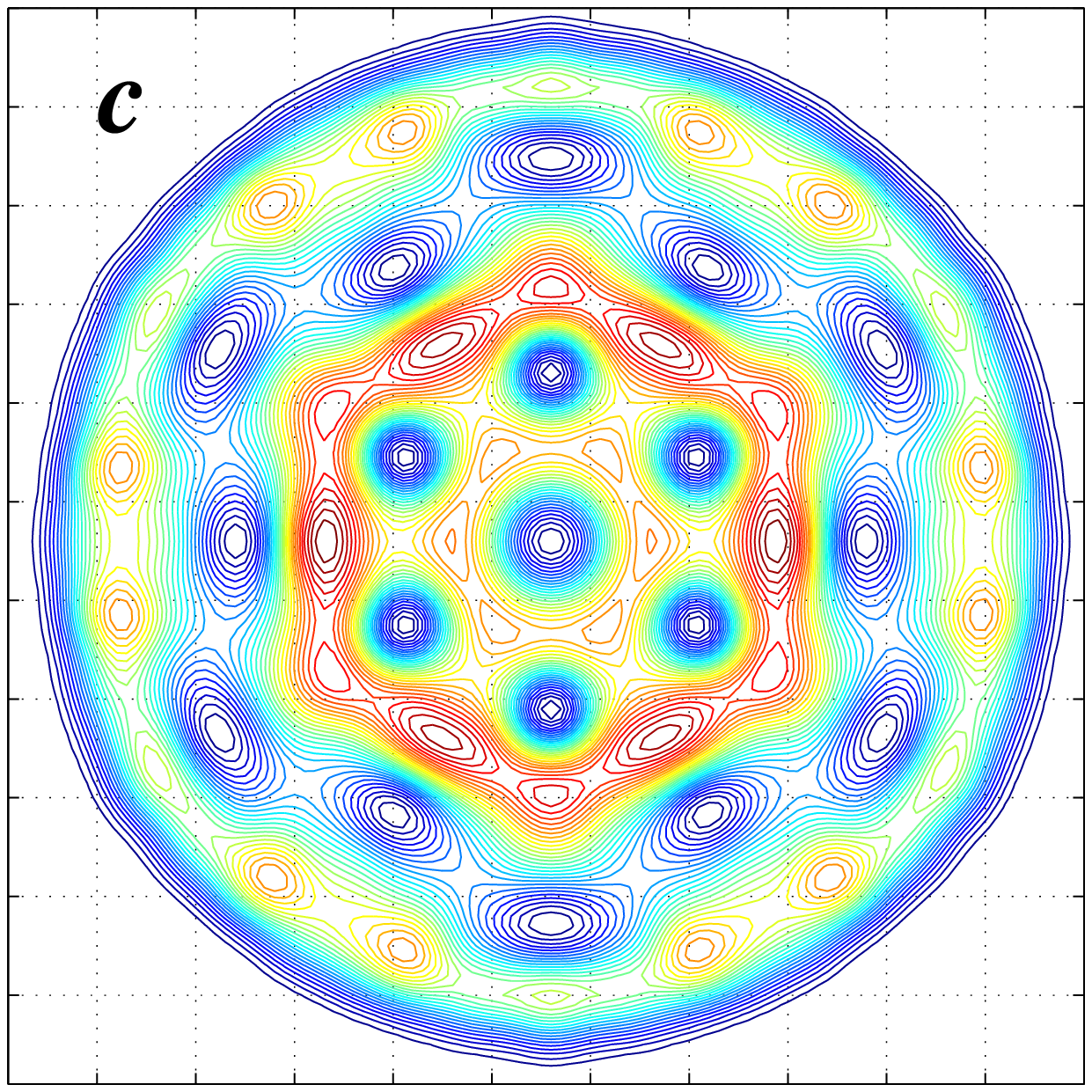}
\end{minipage}
\hspace{-0.5in}
\begin{minipage}[t]{0.4in}
\vspace{0.8in}
\includegraphics[width=0.4in]{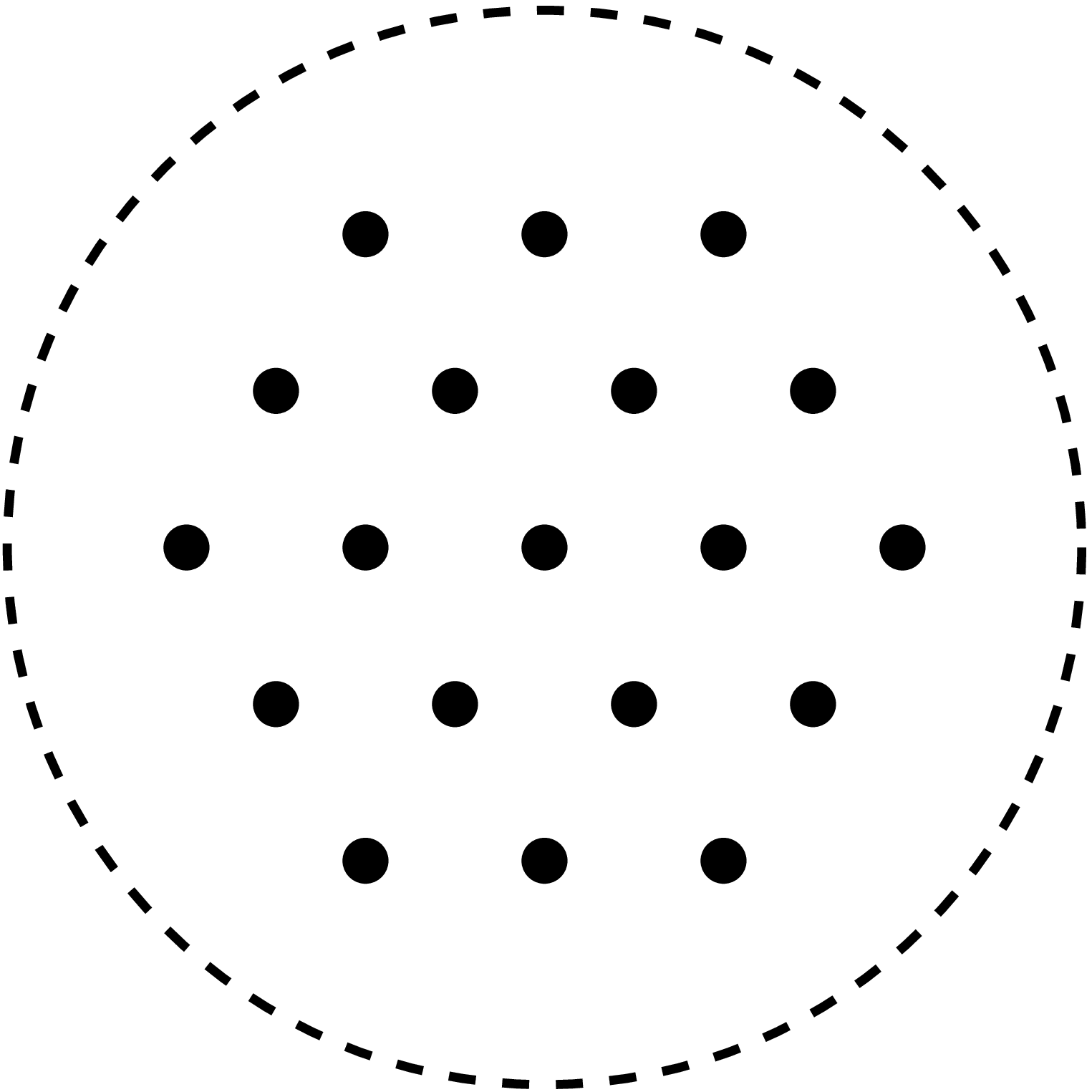}
\end{minipage}
}
\vspace{0.15cm}
\caption[]{
a) PL profile for $2$ vortices placed at $x/R=\pm 0.2$;
b), c) PL profile for hexagonal arrays of 7 and 19 vortices.
The insets show the configuration of vortices in real space.
}
\label{fig:vort_array}
\end{figure}

Remarkably, the property that the pattern of the nodes in momentum distribution 
is a replica of the vortex arrangement, appears to be generic. Although the
replica is exact only for `vortices' of the form
(\ref{eq:prod(z-zj)}) and Gaussian density profiles, 
all the qualitative features hold also 
for the phase vortices
(\ref{eq:phase_vortices}).
As an illustration, we consider a single phase vortex (\ref{eq:phase_vortices})
located some distance $x$ away from trap center. 
As can be seen in Fig. \ref{fig:vort_move},
in this case there is exactly one node in momentum distribution,
shifted off center by a distance that increases with $x$. 
The node displacement orientation
is at a $90^{\circ}$ angle relative to the vortex displacement.
A similar observation can be made for two vortices (Fig.~\ref{fig:vort_array}a).

Also, we consider the configuration of nodes in $N_{\rm ex}(\vec k)$ corresponding 
to a vortex array of the form (\ref{eq:phase_vortices}). Approximately, 
the nodes in this case also form a $90^{\circ}$-rotated array, 
with the distortion being a function of the inter-vortex spacing 
(Fig.~\ref{fig:vort_array}b,c).

Let us now turn to the discussion of finite temperature effects.
At non-zero temperatures, in the presence of density and phase fluctuations, 
$\psi=\sqrt{\rho + \pi}e^{i\phi}$, we construct a perturbation theory
in  $\pi$, $\phi$.
As discussed by Hohenberg and Fisher
\cite{fisher88:_dilut_bose_gas_two_dimen}, 
who considered the perturbation expansion
for weakly non-ideal Bose gas, this is only strictly valid
at very low densities, $\ln(\ln(1/na^2))\gg1$.  Their renormalisation
group analysis however suggests that no other interactions are relevant.

There are three temperature regimes, a low temperature part dominated by
phase fluctuations, a free particle like regime, and a vortex dominated
regime near the transition temperature.  We consider only the phase
fluctuations: it will become clear in a moment that 
the temperature dependence of the PL peak is predominantly due to the phase
fluctuation effects. 
In this case, equation~(\ref{eq:2}) becomes:
\begin{equation}
  \label{eq:6}
  N_{\rm ex}(\vec{k}) 
  =
  \sum_{\vec{r},\vec{r}^{\prime}}
  (\rho(\vec{r}) \rho(\vec{r}^{\prime}))^{1/2}
  e^{-\frac12 D_{\vec r,\vec r'}}
  e^{i \vec{k}(\vec{r}-\vec{r}^{\prime})}
\end{equation}
with $D_{\vec r,\vec r'}=
\left< \left(\phi(\vec{r}) - \phi(\vec{r}^{\prime})\right)^2 \right>$. 
The phase fluctuations are determined by the energy functional 
\begin{equation}
  \label{eq:7}
  H[\phi] = \int \frac{\rho(\vec r)}{2m}\lp\nabla \phi(\vec r)\rp^2 d^2\vec r
\end{equation}
so that the Green's function 
$G(\vec{r},\vec{r}^{\prime})= \left<
  \phi(\vec{r}) \phi(\vec{r}^{\prime}) \right>$  obeys
\begin{equation}
  \label{eq:8}
  - \frac{\beta}{m}
  \nabla \left(
    \rho(\vec{r}) \nabla G(\vec{r},\vec{r}^{\prime}) 
  \right)
  = 
  \delta(\vec{r}-\vec{r}^{\prime})
\end{equation}

For a uniform density profile the solution
is \cite{popov83:_funct_integ_qft} 
\be
  \frac{1}{2}
  \left<
    \left(\phi(\vec r) - \phi(\vec r^{\prime})\right)^2
  \right>
  = 
  \frac{m}{2\pi\beta\rho}
  \ln\left( \frac{|\vec r-\vec r^{\prime}|}{\xi_T} \right) 
\ee
with the thermal length 
$\xi_T=(\lambda\rho/4m)^{1/2}/k_{\rm B}T$ 
corresponding to the energy cutoff in the Planck distribution 
of phase fluctuations. 

One can use this result to estimate 
the fluctuation effect for the trapped gas.
It is convenient to rewrite the logarithm so as to
identify a separation dependent and an equal point term,
\begin{eqnarray}
  \label{eq:9}
  \frac{1}{2}\left(
    G(\vec r^{\prime},\vec r^{\prime}) + G(\vec r,\vec r) - 2 G(\vec r,\vec r^{\prime})
    \right)=
    \nonumber\\
    -\frac{m}{2\pi\beta\rho} \left(
    \ln\left( \frac{\xi_T}{R} \right) -
    \ln\left( \frac{|\vec r-\vec r^{\prime}|}{R} \right) 
  \right)
\end{eqnarray}
We consider
in particular the case where $\vec{r}$ and $\vec{r}^{\prime}$ are well
separated, so $G(\vec{r},\vec{r}^{\prime})$ is small compared to
$G(\vec{r},\vec{r})$. (The typical separation $\vec r-\vec r'$ in 
Eq.(\ref{eq:6}) is of the order of the cloud radius $R$.) In this case
\begin{equation}
  \label{eq:10}
  N_{\rm ex}(\vec{k}) =
  \left| 
    \int 
    \sqrt{\rho(\vec{r})}
      e^{-\frac12 G(\vec{r},\vec{r})}
    e^{i \vec{k}.\vec{r}} d^2 \vec{r}
  \right|^2
\end{equation}
Physically, this result means that the phase fluctuations are dominated by
the equal point contribution, 
which allows one to make a local density approximation 
in Eq.(\ref{eq:9}) and write $G(\vec{r},\vec{r})=(m/2\pi\beta\rho(\vec r))\ln(R/\xi_T)$.


Let us now discuss the phase fluctuations more systematically,
and verify the local density approximation.
We shall consider the condensate 
in an harmonic trap.
Returning to (\ref{eq:8}), rescaling 
$\vec r=R\vec t$, and using $\vec t$ as a new variable, $\rho(t) = \rho_0 (1-t^2)$,
we write the solution to (\ref{eq:8}) as a sum of angular modes,
\begin{equation}
  \label{eq:11}
  G(\vec{t},\vec{t}^{\prime})= -\frac{m}{\beta \rho_0}
  \frac{1}{2\pi}
    \sum_{l=0}^{\infty} g_l(t,t^{\prime}) e^{i l(\theta-\theta^{\prime})}
\end{equation}
%
The equation for mode $l$ has the form:
\begin{displaymath}
  t^2(1-t^2)\frac{d^2g_l}{dt^2} +
  t(1-3t^2)\frac{dg_l}{dt} - l^2(1-t^2)g = t \delta(t-t^{\prime})
\end{displaymath}
The substitutions $g_l(t) = t^{\pm l} f(t)$ and $y=t^2$ show this is
the hypergeometric equation \cite{whittaker27:_moder_analy}, the
general solution is
\begin{equation}
  \label{eq:12}
  g_l(t,t^{\prime}) = t^l_< h_{+l}(t_<) \left[
    t^{l}_> h_{+l}(t_>) - t^{-l}_> h_{-l}(t_>)
    \right]
\end{equation}
where $h$ is an hypergeometric function:
$$
  h_{\pm}(t) = F(a,b,c;t^2)
\,,\quad
  a+b = c = 1 \pm l,\quad  ab=\pm l/2
$$
and $t_<$($t_>$) correspond to the smaller(larger) of $t,t^{\prime}$.


As $t\rightarrow t^{\prime}$, 
the value of $G$ is dominated by the large $l$ terms, 
for which the functions $h_\pm(t)$ tend to
$1/\sqrt{2|l|(1-t^2)}$ and so we can consider the terms:
\begin{displaymath}
  g_l(t+\epsilon,t-\epsilon) = \frac{1}{1-t^2} \frac{1}{2l} 
  \left[t^{2l} \left\{ 1 + 
      {\mathcal{O}}\left(\epsilon^2\right)
    \right\} +
    \left(1 - \frac{2l\epsilon}{t} \right)
  \right]
\end{displaymath}
The second of the terms in brackets gives a divergence as
$\epsilon \rightarrow 0$, of the form
\begin{eqnarray}
  \label{eq:13}
  \lim_{\epsilon \rightarrow 0} 
  G(t+\epsilon,t-\epsilon) 
  &\approx&  \frac{-m}{\pi \beta \rho_0(1-t^2)}  
  \sum_{l=1}^{\infty}
  \frac{1}{2l} \left(
    1 -  \frac{\epsilon}{t}
  \right)^{2l}\nonumber\\
  & =&  \frac{- m}{2\pi\beta \rho(t)}
  \ln\left(
    \frac{2\epsilon}{t}
  \right) 
\end{eqnarray}
This matches both the nature of the divergence, and the pre-factor
($-m/2\pi\beta \rho$) seen in equation~(\ref{eq:9}), and supports the local density approximation (\ref{eq:10}) for a finite system, with $\rho$ replaced by
$\rho(\vec r)$.


From the asymptotes of the trapped gas Green's functions (\ref{eq:13}),
cutting the logarithmic divergence with the thermal length, the
expression for momentum distribution due to far separated excitons
(\ref{eq:10}) becomes:
\begin{equation}
  \label{eq:14}
    N_{ex}(\vec{k}_\parallel ) =
  \left| 
    \int 
    \left(\frac{\lambda\rho(\vec{r})}{m R^2T^2}\right)^{m T/2\pi\rho(\vec{r})}
 \!\!\!   \sqrt{\rho(\vec{r})} e^{i \vec{k} \vec{r}} d^2 \vec{r}
  \right|^2
\end{equation}
This expression has the form of replacing the condensate density with
a ``coherent particle density'',
\begin{equation}
  \label{eq:15}
  \rho(\vec{r}) \rightarrow
  \rho(\vec{r})
  \left(\frac{\rho(\vec{r})}{\rho_B}\right)^{\rho_A/\rho(\vec{r})}\\
\end{equation}
For an harmonic trap, with the numbers discussed above,
\begin{eqnarray*}
&&  \frac{\rho_A}{\rho_0} 
  = \frac{m k_{\rm B} T}{8 \pi \hbar^2}\sqrt{\frac{\pi \lambda}{2 \alpha N}}
    \sim 
  30 \times\frac{T [{\rm K}]}{\sqrt{N}}
\\
&&  \frac{\rho_0}{\rho_B} 
  =
  \frac{ \alpha 2\pi\hbar^2}{m (k_{\rm B} T)^2}
  \sim 
  \left( \frac{10^{-2}}{T [{\rm K}]} \right)^2
\end{eqnarray*}
The temperature dependence of the PL peak is shown in Fig.\ref{fig:peak}.
Qualitatively,
the peak is suppressed at 
\be
  \label{eq:16}
  T > T_* = T_{\rm BEC} / \ln(R/\xi_T)
\ee
i.e. well below the condensation transition.

In summary, we propose that the angular profile of PL can be used as a
diagnostic of Bose condensation in 2D exciton traps. 
A peak in the emission in a direction normal to the 2D plane appears 
due to long-range phase coherence in the condensate, at temperatures
for which the coherence length exceeds the wavelength of the emitted radiation.
The distribution of PL radiation in the peak is sensitive to the presence 
of phase textures in the condensate. In particular, vortices are imaged 
by the nodes in the PL angular profile.

We are grateful to L.V.\,Butov and B.D.\,Simons for useful
discussions, and also for support from the Cambridge-MIT institute and
the EU network ``Photon-Mediated phenomena in semiconductor
nanostructures'' HPRN-CT-2002-00298.  The NHMFL is supported by the
National Science Foundation, the state of Florida and the US
Department of Energy.

\end{multicols}

\end{document}